\documentclass[10pt]{article}

\usepackage{times,mathptmx}

\usepackage{geometry}
\usepackage{amsmath,amsthm,amssymb}

\usepackage[font=small]{caption}

\usepackage{graphics}
\usepackage{graphicx}
\usepackage{color}

\usepackage[compact]{titlesec}
\titleformat{\section}{\large\bfseries}{\thesection}{1em}{}
\titleformat{\subsection}{\normalsize\bfseries}{\thesubsection}{1em}{}
\titlespacing*{\section}{0em}{1em}{0.2em}
\titlespacing*{\subsection}{0em}{0.8em}{0.2em}
\titlespacing*{\subsubsection}{0em}{0.8em}{0.2em}

\usepackage{fancyhdr}
\usepackage[square,semicolon]{natbib}
\renewcommand{\cite}[1]{\citep{#1}} % make cite alias citep
\usepackage[T1]{fontenc}   % for umlauts and other diaeresis
\usepackage[utf8]{inputenc} % for a UTF8 editor only

\makeatletter
\renewcommand{\@biblabel}[1]{}
\makeatother

\renewcommand{\title}[1]{
  \begin{center}
    {\LARGE \textbf{#1}}
  \end{center}
}

\makeatletter
\newcommand{\authorinfo}[2]{
  \ifdefined \authorinfos
    \protected@edef \authorinfos{\authorinfos; \textsuperscript{#2}#1}
  \else
    \protected@edef \authorinfos{\textsuperscript{#2}#1}
  \fi
}
\renewcommand{\author}[2]{
  \ifdefined \authors
    \protected@edef \authors{\authors, #1\textsuperscript{#2}}
  \else
    \protected@edef \authors{#1\textsuperscript{#2}}
  \fi
}
\makeatother
\newcommand{\printauthors}{
  \begin{center}
    \noindent {\large \textbf{\authors}} \\
    \textit{\authorinfos}
  \end{center}
}

\newcommand{\correspondence}[2]{
  \begin{center}
    {\small Correspondence: #1. E-mail:~#2}
  \end{center}
}

\renewcommand{\abstract}[2]{
  \hrule
  \vspace{0.6em}
  \noindent \textbf{\textit{Abstract.}} #1 \par
  \vspace{1em}
  \noindent {\small\textit{Keywords:} #2}
  \vspace{0.6em}
  \hrule
  \vspace{2em}
}

\begin{document}

\title{Evaluation of a congruent auditory feedback for Motor Imagery BCI}

% enter author information here
% first argument is author info, institution, city, state, country
% second argument is superscript, make sure the matches an author below
\authorinfo{PRISM laboratory, CNRS, Univ. Aix Marseille, France}{1}
\authorinfo{Ullo, La Rochelle, France}{2}
\authorinfo{Clinique du Sommeil, CHU de Bordeaux, CNRS, Univ. Bordeaux, France}{3}
\authorinfo{Inria, Bordeaux, France}{4}
\authorinfo{CRNL Inserm, CNRS, Lyon, France}{5}
\authorinfo{ENS, Pays germaniques Archives Husserl,CNRS, Paris, France}{6}

% enter author names here
% first argument is author name
% second argument is superscript, match to author info above
%   multiple author superscripts may match a single author info
\author{Emmanuel~Christophe}{1}
\author{Jérémy~Frey}{2}
\author{Richard~Kronland-Martinet}{1}
\author{Jean-Arthur~Micoulaud-Franchi}{3}
\author{Jelena~Mladenović}{4,5}
\author{Gaelle~Mougin}{6}
\author{Jean~Vion-Dury}{1}
\author{Solvi~Ystad}{1}
\author{Mitsuko~Aramaki}{1}

\printauthors

\correspondence{Jelena Mladenović, Inria, Bordeaux, France}{jelena.mladenovic@inria.fr}

\abstract{Designing a feedback that helps participants to achieve higher performances is an important concern in brain-computer interface (BCI) research. In a pilot study, we demonstrate how a congruent auditory feedback could improve classification in a electroencephalography (EEG) motor imagery BCI. This is a promising result for creating alternate feedback modality.}
{BCI, EEG, motor imagery, auditory feedback, congruent feedback}

\section{Introduction}

Designing a feedback that helps participants to achieve higher performances is an important concern in  brain-computer interface  (BCI) research. Various congruent (task-related) visual feedback have been examined and showed promising results, e.g. using body ownership illusions in VR \cite{alimardani2014effect}. The use of congruent sound in combination with the visual modality demonstrated to increase performances in Motor Imagery (MI) BCI \cite{tidoni2014audio}. However, evaluating solely auditory feedback is often neglected, even though it could be a valuable alternative modality. We investigate the relevance of congruent versus non-congruent (abstract) auditory feedback in assisting the user to imagine feet movement in an EEG-based MI BCI.

\section{Material, Methods and Results}

\begin{figure}[h!]
  \vspace{-1em} % ditch extra space above figures
  \begin{center}
    \includegraphics[width=0.90\textwidth]{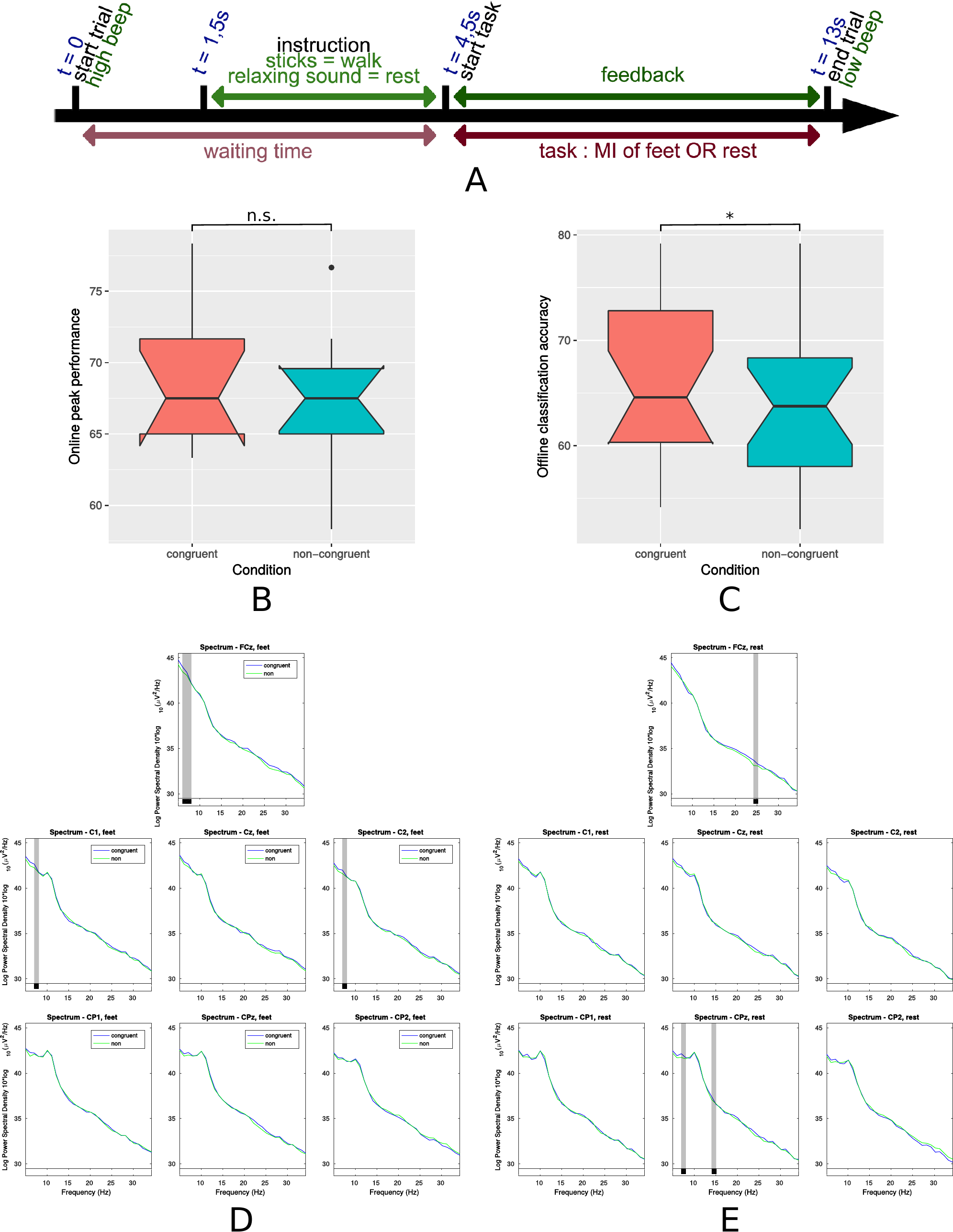}
    \vspace{-1em} % ditch space above caption
    \caption{\textit{Comparison between congruent and non-congruent audio feedback during a motor imagery BCI (feet imagination vs rest). \textbf{A}: Audio sequence for one trial. \textbf{B}: Peak performance of the online classifier. \textbf{C}: Accuracy of a classifier trained offline on each conditon. \textbf{D} and \textbf{E}: spectral power analysis per elecrtode, during feet and rest trials respectively (significant differences are highlighted, $p < 0.05$).}}
    \label{fig:plot}
  \end{center}
  \vspace{-1.5em} % ditch extra space below figures
\end{figure}

Users task was either to imagine moving their feet or to rest. The instructions consisted of 4-taps of sticks for MI of feet, and a relaxing sound for the resting state (Figure \ref{fig:plot} A). There were two conditions: "non-congruent", during which the feedback provided was abstract and not related to any task (harmonic sounds with a different pitch) and "congruent", during which the MI feedback reflected the sound of one’s footsteps on gravel, and the rest feedback a relaxing sound of water. The choice of such sound for the congruent MI task was motivated by the fact that rhythmical sounds relate to motor cortex \cite{bengtsson2009listening}. For the same reason, both tasks in the non-congruent condition were as non-rhythmic as possible. Moreover, such congruent sounds were chosen to increase sense of ownership. The influence of the feedback was evaluated within-subjects.

Non-congruent and congruent feedback were generated in real-time, the former with Max-MSP and the latter with a dedicated synthesizer of environmental sounds \cite{verron20103}.

Ten participants were recruited (2 women, mean age: 24.8, SD: 4.98, all BCI naives). They were seated in front of a single speaker. 7 passive gold cup electrodes were placed over Cz, C1, C2, FCz, CPz, CP1 and CP2 in the 10-20 system and connected to an OpenBCI Cyton amplifier. Before the experiment participants heard an example of each sound to get accustomed to the task. In order not to reveal the outcome of the experiment, the feedback was simply introduced as “environmental sounds” or “musical sounds”. A run of calibration was followed by two runs with one feedback and then two runs with the other (conditions' order was counterbalanced across participants). A run contained 30 trials of 13 seconds, 15 for each class (rest or MI) in random order. A session lasted $\approx$45m.

Using OpenViBE, signals were filtered in the alpha (8-13 Hz) and beta (13-30 Hz) bands, passed to a Filter Bank Common Spatial Pattern filter and classified using linear discriminant analysis. We tested for significance using Wilcoxon signed-rank tests. There was no difference in online performance between the two feedback (Figure \ref{fig:plot} B). An offline analysis revealed a significant difference (p < 0.05) in classification accuracy when a classifier was trained separately on the "congruent" and "non-congruent" runs (respectively 66.1\%, SD: 7.45 and  63.9\% SD: 7.8, 10-fold cross-validation, Figure \ref{fig:plot} C). There were also changes in EEG spectral power, with more activation in the "congruent" condition within the beta band during rest and within both alpha and beta bands during MI (Figure \ref{fig:plot} D and \ref{fig:plot} E).

\section{Discussion}

While the online performance remained unchanged with a congruent feedback, an offline classifier could benefit from increased differences between rest and MI signals. Additionally, post-hoc interviews revealed that participants felt assisted by a congruent feedback.

\section{Significance}

In a pilot study, we demonstrate how a congruent auditory feedback could improve classification in a EEG MI BCI, a promising result for creating alternate feedback modality. This prompts for further investigations on a larger sample and with more channels to better assess the underlying change in brain activity.

%%%%%%%%%%%%%%%%%%%%%%%%%%%%%%%%%%%%%%%%%%%%%%%%%%%%%%%%%%%%%%%%%%%%%%

% change template_BCI-meeting_2013_USletter to the name of your bibtex file
%{\footnotesize \bibliography{biblio}}
{\bibliography{biblio}}

\begin{thebibliography}{}

\bibitem[Alimardani et~al., 2014]{alimardani2014effect}
Alimardani, M., Nishio, S., and Ishiguro, H. (2014).
\newblock Effect of biased feedback on motor imagery learning in
  bci-teleoperation system.
\newblock {\em Frontiers in systems neuroscience}, 8:52.

\bibitem[Bengtsson et~al., 2009]{bengtsson2009listening}
Bengtsson, S.~L., Ullen, F., Ehrsson, H.~H., Hashimoto, T., Kito, T., Naito,
  E., Forssberg, H., and Sadato, N. (2009).
\newblock Listening to rhythms activates motor and premotor cortices.
\newblock {\em cortex}, 45(1):62--71.

\bibitem[Tidoni et~al., 2014]{tidoni2014audio}
Tidoni, E., Gergondet, P., Kheddar, A., and Aglioti, S.~M. (2014).
\newblock Audio-visual feedback improves the bci performance in the
  navigational control of a humanoid robot.
\newblock {\em Frontiers in neurorobotics}, 8:20.

\bibitem[Verron et~al., 2010]{verron20103}
Verron, C., Aramaki, M., Kronland-Martinet, R., and Pallone, G. (2010).
\newblock A 3-d immersive synthesizer for environmental sounds.
\newblock {\em IEEE Transactions on Audio, Speech, and Language Processing},
  18(6):1550--1561.

\end{thebibliography}

%%%%%%%%%%%%%%%%%%%%%%%%%%%%%%%%%%%%%%%%%%%%%%%%%%%%%%%%%%%%%%%%%%%%%%
\end{document}